\documentstyle[emulateapj]{article}
\def\keywords{}
\def\acknowledgements{}
\def\apj{ApJ}
\def\aj{A.J.}

\def\apjs{ApJS}

\begin{document}

\def\nth{n_{\rm th}}
\def\nobs{n_{\rm obs}}
\def\dmin{d_{\rm min}}
\def\macho{{\sc macho}}
\def\newpage{\vfill\eject}
\def\vs{\vskip 0.2truein}
\def\gnu{\Gamma_\nu}
\def\fnu {{\cal F_\nu}}
\def\mass{m}
\def\lum{{\cal L}}
\def\imf{\xi(\mass)}
\def\ilf{\psi(M)}
\def\msun{M_\odot}
\def\zsun{Z_\odot}
\def\met{[M/H]}
\def\vi{(V-I)}
\def\mtot{M_{\rm tot}}
\def\mhalo{M_{\rm halo}}
\def\pp{\parshape 2 0.0truecm 16.25truecm 2truecm 14.25truecm}
\def\la{\mathrel{\mathpalette\fun <}}
\def\ga{\mathrel{\mathpalette\fun >}}
\def\fun#1#2{\lower3.6pt\vbox{\baselineskip0pt\lineskip.9pt
  \ialign{$\mathsurround=0pt#1\hfil##\hfil$\crcr#2\crcr\sim\crcr}}}
\def\ie{{ i.e., }}
\def\eg{{ e.g., }}
\def\etal{{et al.\ }}
\def\etalc{{et al., }}
\def\kpc{{\rm kpc}}
 \def\Mpc{{\rm Mpc}}
\def\mh{\mass_{\rm H}}
\def\mmax{\mass_{\rm u}}
\def\ml{\mass_{\rm l}}
\def\bc{f_{\rm cmpct}}
\def\br{f_{\rm rd}}
\def\kmsec{{\rm km/sec}}
\def\ibl{{\cal I}(b,l)}
\def\dmax{d_{\rm max}}
\def\dmin{d_{\rm min}}
\def\mbol{M_{\rm bol}}
\def\kms{{\rm km}\,{\rm s}^{-1}}

\lefthead{Graff \& Gould}
\righthead{High Velocity Star Formation in the LMC}
%%%%%%%%%%%%%%%%%%%%%%%
%%%%%%%electronic submission format
\submitted{Submitted to ApJL, 14 February 2000, Accepted 4 March 2000}
\title{High Velocity Star Formation in the LMC}
\author{David S. Graff and Andrew P. Gould}
\affil{Departments of Astronomy and Physics, The Ohio State University,
Columbus, OH 43210, USA}
\authoremail{graff.25@osu.edu, gould@payne.mps.ohio-state.edu}
%%%%%%%%%%%%%%%%%%%%%%%%%%%%%%%

\begin{abstract}
Light-echo measurements show that SN1987A is 425 pc behind the LMC
disk.  It is continuing to move away from the disk at $18\,\kms$.
Thus, it has been suggested that SN1987A was ejected from the LMC
disk.  However, SN1987A is a member of a star cluster, so this entire
cluster would have to have been ejected from the disk.  We show that the
cluster was formed in the LMC disk, with a velocity perpendicular to
the disk of about $50\,\kms$.  Such high velocity formation of a star cluster is unusual,
having no known counterpart in the Milky Way.
\end{abstract}

\keywords{Magellanic Clouds, Supernovae: Individual (SN1987A)}
%%%%%%%%%%%%%%%%%%%%%%%%%%%%

\setcounter{footnote}{0}
\renewcommand{\thefootnote}{\arabic{footnote}}

\section{Introduction}

The Large Magellanic Cloud (LMC) shows a clear contrast between
regular kinematics and irregular structure, with its offcenter bar and
lack of any clear stellar spiral morphology.  The velocities as traced
by carbon star velocities (Graff \etal 2000; Hardy \etal 2000; Kunkel
\etal 1997) and by H$\alpha$ emmision (Kim \etal 1999) are well fit by
a rotating disk although there may be a non-disk component (Graff
\etal 2000; Luks \& Rohlfs, 1992).  The overall velocity dispersion of
the carbon stars $\sim 20\,\kms$ is small compared to the rotational
velocity of the LMC $(60 - 70 \,\kms)$ indicating that the stellar
component of the LMC is relatively flat and rotationally supported.
Moreover, Graff \etal (2000) showed that the younger, metal rich
carbon stars in the inner $4^\circ$ of the LMC have a much lower
velocity dispersion, only $8\,\kms$.  This contrast suggests that
the LMC lies in a nearly face-on plane, but is irregular within
that plane.

The three-dimensional structure of LMC dust was measured using
the ``light-echo'' technique on SN1987A by Xu, Crotts \& Kunkel
(1996).  They identified 12 seperate dust sheets.  Most significantly, in this
work and in a follow up spectroscopic study (Xu \& Crotts 1999), they
identified three components with the spherical shell N157C enclosing the OB 
association LH 90.  This shell was found to lie 490 pc in
front of SN1987A.  Including the LMC inclination of $\sim 30^\circ$, 
the component of this distance perpendicular to the LMC plane
is 425 pc.

This distance is much greater than the virial thickness of the young
stellar population of the LMC in the location of SN1987A $\sim 90$ pc
(given the local surface density of 100 $\msun/{\rm pc}^2$ which we
determine below).  Thus, it is difficult to imagine how these two
young stellar populations came to be so separated.  Xu \etal (1996)
suggest that SN1987A is a ``...runaway star behind the disk of the
Large Magellanic Cloud''.

Classical runaway stars can be found high above the Milky Way plane
(Conlon \etal 1990).  The runaway O and B stars are thought to be
ejected by one of two processes: supernova explosions in close binary
systems (Blaauw 1961) and strong dynamical interactions in star
clusters (Poveda, Ruiz, \& Allen 1967; Gies \& Bolton 1986).  Indeed, Hipparcos
measurements of O and B stars have found several runaways that can be
identified as having been ejected from particular OB associations (de
Zeeuw \etal 1999).

However, Efremov (1991) and Panagia \etal (2000) have identified
SN1987A as belonging to KMK 80 (Kontizas, Metaxa \& Kontizas 1988)
``...a loose young cluster $12 \pm 2$ Myr old....'' (Panagia \etal
2000).  Thus, it cannot be a classic runaway star; any of the violent
ejection mechanisms discussed above would eject only the single star,
and not its cluster.

 In the next section, we solve for the initial kinematics of SN1987A
and its associated cluster, KMK 80.  We find that the cluster {\it
formed} in the LMC plane, moving with a velocity of $50\,\kms$
perpendicular to the LMC plane.

\section{Kinematics of SN1987A}

We begin by examining the velocities of these two young
clusters relative to the LMC.  The disk solution of Hardy et al.\
(2000) at the projected position of SN1987A is $271 \pm 1\,\kms$.
% (with the nearest clump of 23
%measured carbon stars -- field 30 of Blanco \& McCarthy 1990 -- 
%redshifted by $5\,\kms$ relative to the disk solution, i.e.,
%$272\pm 3\,\kms$).

  By comparison, SN1987A has a redshift of $286\,\kms$ (Meaburn,
Bryce \& Holloway 1995) while the N157C complex containing LH 90 has a
redshift of $270\,\kms$ (Xu \& Crotts 1999).  Thus, the velocity of LH 90
is perfectly consistent with the LMC disk velocity at this point.

On the other hand, SN1987A is in two respects inconsistent with being
a member of the cold population: first, it is moving $15\,\kms$
relative to the disk, faster than the $8\,\kms$ typical of the cold
population.  Secondly, and more importantly, it lies far above the
scale height of the cold population (and even above the scale height
of the hot population).

To take account of both effects simultaneously, we define the
``verticle energy'' of a star to be $E \equiv v_z^2/2 + \Phi(z)$ and
approximate the potential energy to be $\Phi(z) \approx 2 \pi G \Sigma
|z|$ for stars of height $z\gg 150\,$pc.  An examination of the
isophotal map of the LMC of de Vaucouleurs (1957) shows that the
surface brightness of the LMC in the neigborhood of SN1987A is about
21.7 mag./arcsec$^2$, or 56 $L_\odot {\rm pc}^{-2}$.  Assigning a
Population I mass-luminosity ratio of 1.7, we derive a mass surface
density of roughly $\Sigma_{\rm SN1987A} \approx 100 M_\odot {\rm
pc}^{-2}$.  We derive a total energy of $1300 \,(\kms)^2$
corresponding to a midplane velocity of $50\,\kms$.  Thus, the total
gravitational energy of the supernova is much too high for it to be a
member of the cold population (and somewhat high even for the hot
population).
 
We note that the age of the star cluster containing the supernova is
about 12 Myr which is consitent with estimates of the age of the
precursor to the supernova.  If the star cluster was formed at the LMC
plane 12 Myr ago, with a velocity perpendicular to the plane of $50 \
\kms$, and this velocity decreased with a gravitational acceleration
of $-3\,\kms\,\rm Myr^{-1}$, it would today be $\sim 400\,$pc above
the plane moving at $14 \ \kms$, consitent with its measured distance
of 425 pc above the plane and relative velocity of $15 \ kms$.

\section{Discussion}

The match between these numbers is compelling, and we suggest that the
entire KMK 80 star cluster was formed 12 Myr ago at the LMC plane, but
with an extraordinarily high velocity of $50\,\kms$ perpendicular to the 
plane.  The agreement between age and flight time is typical of most runaway 
O and B stars in the halo of the Milky Way (Keenan, Brown \& Lennon 1986).

We do not know what mechanism could create a star cluster moving at
such high velocities.  As far as we know, there is no counterpart in
the Milky Way.  However, we can speculate on two possible mechanisms.
First, the cluster might have formed as part of a galactic fountain
pushed out of the LMC by supernovae or stellar winds.  Such a
mechanism was put forward by Xu \& Crotts who suggested that SN1987A
was formed on a shell of gas pushed out of the LMC by LH 90.  These
authors noted that SN1987A is on the outskirts of the extremely
violent 30 Dor. region. 

Second, a dense cloud of gas could have smashed through the LMC disk,
triggering star formation in the process with the resulting stars
carrying some of the initial momentum of the cloud.  This cloud could
have been fountain material raining back down onto the LMC disk, or it
could have been a high velocity cloud orbiting either the LMC or the
Milky Way.

There are a few systems in the Milky Way that might have been formed
in processes similar to the KMK 80 cluster.  In addition to runaway O
stars, the Milky Way Halo also contains young, high velocity, high
metallicity A stars (Perry 1969; Rodgers 1971).  These stars are all
roughly the same age, $<650\,$Myr (Lance 1988), which suggests that
they were created from the collsion of a Magellanic Cloud sized galaxy
with the Milky Way disk (Rodgers, Harding, \& Sadler 1981; Lance
1988).  A similar recent collision in the LMC might generate high
velocity star formation without breaking up KMK 80.

Gould's belt (Gould 1874; P\"oppel 1997) contains several OB
associations in a roughly planar region oriented $18^\circ$ from the
plane of the Milky Way.  Comer\'on \& Torra (1994) suggested that
Gould's belt arose from the glancing collision of a high velocity
cloud with the Milky Way disk.  Perhaps KMK 80 is part of a
similar structure oriented more nearly perpendicular to the LMC plane.

Logically, there are only two alternatives to our interpretation that
KMK 80 formed at high vertical velocity.  First, KMK 80 may actually
lie in the LMC plane while the progenitor of SN1987A is simply seen
projected against this cluster, having been earlier ejected from a
binary.  This appears to us to be a priori unlikely and can in any
event be tested by spectroscopic observations of KMK 80 members.  In
addition to confirming SN1987A as a radial-velocity member of this
cluster, such measurements would yield the metallicity of the cluster
and so of the SN1987A progenitor.

Second, SN1987A could actually lie in the LMC plane while N157C lies
490 pc closer to us.  Then, either LH 90 would still be at the center
of N157C, or it would lie in the LMC plane and be seen by chance
projected against the center of this cloud.  In the first case, one
would still have the same problem of an OB association lying far from
the LMC plane.  As for the second case, the probability of a chance
projection of two such naturally associated structures seems
incredibly low.  In either case, KMK 80 would have to have
been born with a vertical energy at least equal to its present
kinetical energy of $(15\,\kms)^2$, which is still quite high.
Moveover, the N157C cloud would have to have exactly the same radial
velocity as the LMC plane despite the fact that it lies $\sim 400\,$pc
from it.  Hence, the various alternatives to our interpretation, while
not actually ruled out, require extraordianry combinations of
coincidences.

\acknowledgements

We thank Arlin Crotts, Yuri Efremov and David Weinberg for useful discussions.
This work was supported in part by grant AST 97-27520 from the NSF.

\end{document}